\documentclass{nature}

\usepackage[super]{natbib} 
\usepackage{array}
\usepackage{times}
\usepackage{graphicx,color}
\makeatletter
\let\saved@includegraphics\includegraphics
\AtBeginDocument{\let\includegraphics\saved@includegraphics}
\renewenvironment*{figure}{\@float{figure}}{\end@float}
\makeatother

\def\arcsec{$^{\prime\prime}$}

\newcommand{\be}{\begin{equation}}
\newcommand{\ee}{\end{equation}}

\newcommand{\kms}{\,{\rm {km\, s^{-1}}}}
\newcommand{\msun}{{$M_{\odot}$}}

\newcommand{\gtsima}{$\; \buildrel > \over \sim \;$}
\newcommand{\ltsima}{$\; \buildrel < \over \sim \;$}
\newcommand{\prosima}{$\; \buildrel \propto \over \sim \;$}
\newcommand{\gsim}{\lower.5ex\hbox{\gtsima}}
\newcommand{\lsim}{\lower.5ex\hbox{\ltsima}}
\newcommand{\simgt}{\lower.5ex\hbox{\gtsima}}
\newcommand{\simlt}{\lower.5ex\hbox{\ltsima}}
\newcommand{\simpr}{\lower.5ex\hbox{\prosima}}

\newcommand{\al}{${\alpha}$}

\title{A 10,000-solar-mass black hole in the nucleus of a bulgeless dwarf galaxy}

\author{Jong-Hak Woo$^{1}$, Hojin Cho$^{1}$, Elena Gallo$^{2}$, Edmund Hodges-Kluck$^{2,3,4}$,
Huynh Anh Le$^{1}$, Jaejin Shin$^{1}$, Donghoon Son$^{1}$,
\& John C. Horst$^{5}$
}

\begin{document}

\maketitle

\begin{affiliations}
 \item Department of Physics \& Astronomy, Seoul National University, Seoul, 08826, Republic of Korea
 \item Department of Astronomy, University of Michigan, Ann Arbor, MI 48109, USA
 \item Department of Astronomy, University of Maryland, College Park, MD 20742, USA
 \item NASA/GSFC, Code 662, Greenbelt, MD 20771, USA
 \item San Diego State University, San Diego, CA 92182, USA
\end{affiliations}

\begin{abstract}
The motions of gas and stars in the nuclei of nearby large galaxies have demonstrated that massive black holes are common\cite{Kormendy&Ho13} and that their masses strongly correlate with the stellar velocity dispersion $\sigma_{\star}$ of the bulge\cite{Ferrarese&Merritt00,Gebhardt+00,Onken+04}. This correlation suggests that massive black holes and galaxies influence each other's growth\cite{Silk&Rees98,King03,Fabian12}. Dynamical measurements are less reliable when the sphere of influence is unresolved, thus it remains unknown whether this correlation exists in galaxies much smaller than the Milky Way, as well as what fraction of these galaxies have central black holes. Light echoes from photoionized clouds around accreting black holes\cite{Blandford&McKee82,Peterson93}, in combination with the velocity of these clouds, yield a direct mass measurement that circumvents this difficulty.
Here we report an exceptionally low reverberation delay of $83\pm14$ minutes between variability in the accretion disk and high velocity H$\alpha$ emission from the nucleus of the bulgeless dwarf galaxy NGC~4395. Combined with the H$\alpha$ line-of-sight velocity dispersion $\sigma_{\rm line}=426\pm1$~km~s$^{-1}$, this lag determines a mass of about 10,000~$M_{\odot}$ for the black hole. 
This mass is among the smallest central black hole masses reported, near the low end of expected masses for heavy ``seeds''\cite{Begelman+06,Djkstra+08,Agarwal+12}, and the best direct mass measurement for a galaxy of this size. Despite the lack of a bulge, NGC~4395 is consistent with the $M_{\rm BH} - \sigma_{\star}$ relation when $\sigma_{\star}$ is measured from the central region. This indicates that the relation need not originate from hierarchical galaxy assembly nor from black hole feedback. 
\end{abstract}

The dwarf galaxy NGC~4395 ($M_{\star} \approx 8\times 10^8 M_{\odot}$) is the least luminous active galactic nucleus (AGN) known to date\cite{Filippenko&Sargent89}, with a monochromatic luminosity at 5100\AA\ $L_{5100} < 10^{40}$~erg~s$^{-1}$. Prior efforts to determine the black hole mass indicate that it hosts an intermediate-mass black hole with $M_{\rm BH}$ in the range $10^4 - 5\times 10^5 M_{\odot}$\cite{Filippenko&Ho03,Edri+12,LaFranca+15,denBrok+15,Peterson+05}. The wide range arises both because of the difficulty of dynamical measurements, which cannot resolve the sphere of influence of the black hole amid a more massive nuclear star cluster\cite{Filippenko&Ho03}, and because of the lack of precise reverberation delay measurements. Nevertheless, previous measurements find short delays between 0.75--3.6~hr\cite{Peterson+05,Desroches+06,Edri+12} that implicate a low-mass nuclear black hole. 

On 2017 May 2 and On 2018 April 8, we measured the $V$-band-to-H$\alpha$ time delay from coordinated observations of NGC~4395 with multiple ground-based instruments (see Methods). These included the Templeton camera on the MDM 1.3m telescope, the MDM4K camera on the MDM 2.4m telescope, the e2v 4K camera on the Lemmonsan Optical Astronomy Observatory (LOAO) 1.0m telescope in Arizona, and the e2V 2K camera on the Mt. Laguna Observatory (MLO) 1.0m telescope in California, where we acquired a series of short exposures in $V$ (MDM 1.3m, LOAO 1.0m, and MLO 1.0m) and H$\alpha$ (MDM 2.4m) over most of each night. 

\begin{figure}
    \centering
    \includegraphics[width=0.49\textwidth]{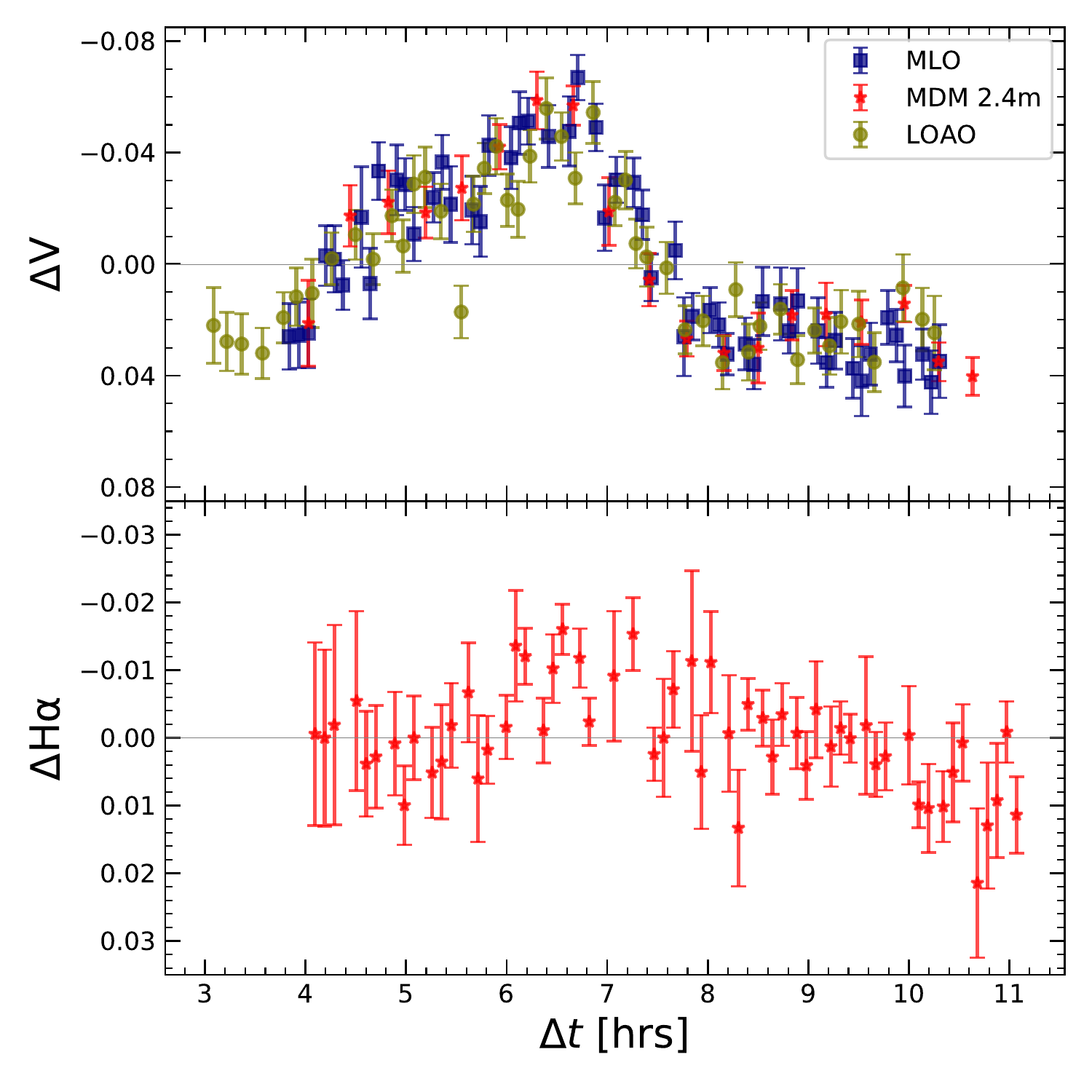}
    \includegraphics[width=0.49\textwidth]{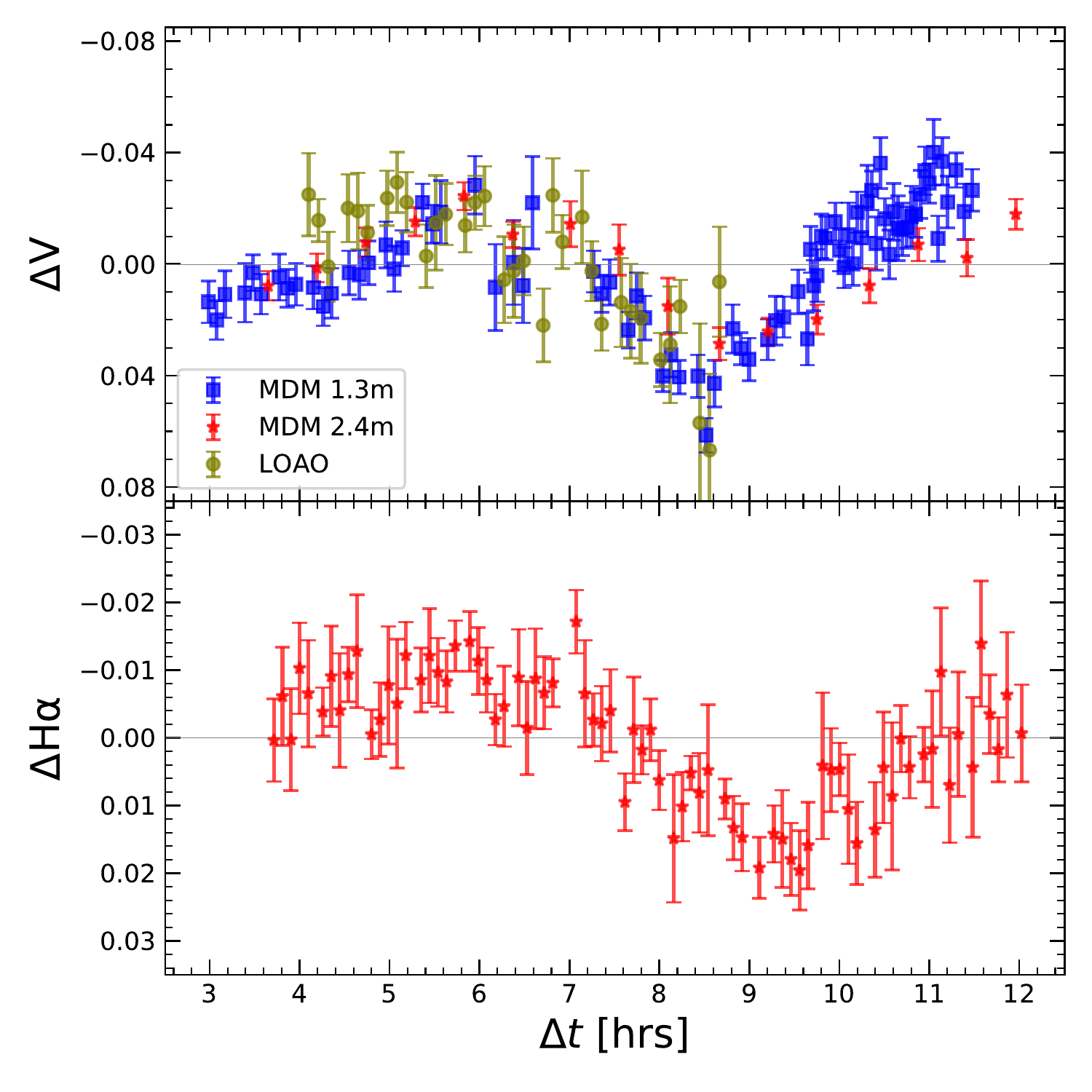}
    \caption{\textbf{$V$- and narrow H$\alpha$-band photometric light curves.} The light curves are generated from the series of images acquired, respectively, with the $V$ (top) and the narrow H$\alpha$ filters (bottom), at MDM [red triangles for the 2.4-m telescope and blue squares for the 1.3-m telescope], LOAO [yellow circles], and MLO [dark blue squares] for the 2017 May 2 (left) and 2018 April 8 (right) campaigns. The $V$-band continuum emission originates in the accretion disk while the H$\alpha$ line flux is emitted from the broad-line region. Thus, the lack between the two light curves reflects the light-travel time.}
    \label{fig:LC}
\end{figure}

\begin{figure}
    \centering
    \includegraphics{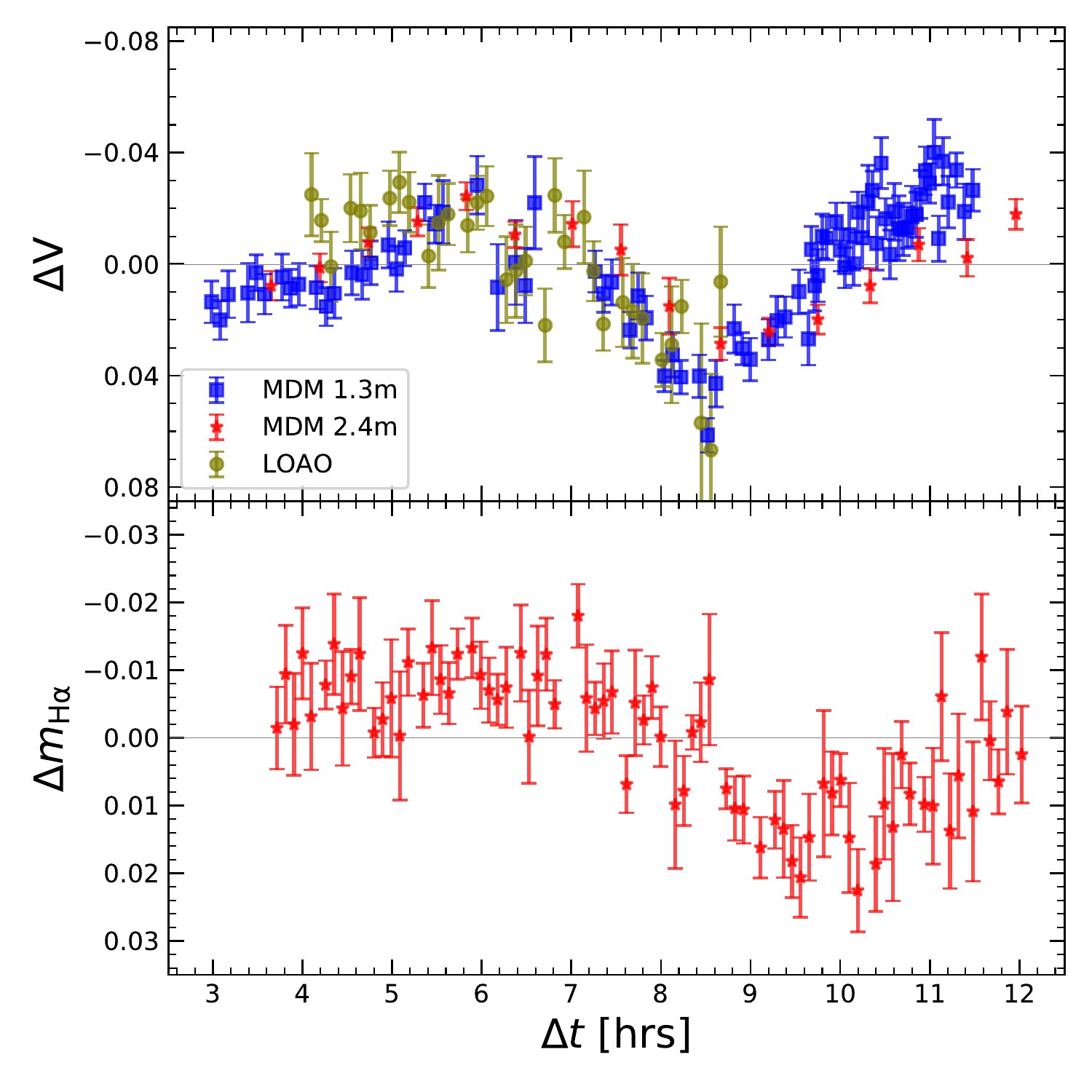}
    \caption{\textbf{Continuum-corrected H$\alpha$ light curve.} The 2018 April 8 H$\alpha$ light curve (bottom) after subtracting the continuum contribution to the narrow H$\alpha$ filter, assuming that the variability is the same as that in the $V$ band (top). The $V$ band data are from the MDM 1.3-m (blue squares), MDM 2.4-m (red triangles), and LOAO 1.0-m (yellow circles) telescopes. As expected, this subtraction leads to a longer lag of $\tau_{\rm cent} = 83\pm14$~minute.}
    \label{fig:LC_corrected}
\end{figure}

A cross-correlation analysis (see Methods) of the photometric light curves (Figure~\ref{fig:LC}) from the higher quality 2018 data set yields a time delay of $\tau = 49^{+14}_{-15}$~minutes. The corresponding $\tau = 55^{+28}_{-32}$~minutes from 2017 is consistent with the 2018 value. This measurement clearly demonstrates the existence of a light-travel time delay between the accretion disk and the broad-line region (BLR), but it is a lower limit because it does not account for the continuum emission in the H$\alpha$ filter, which comes from the accretion disk itself. To recover the true delay, we subtract the maximum continuum contribution by assuming that the continuum variability in the H$\alpha$ band is the same as that measured in the $V$ band (Figure~\ref{fig:LC_corrected}). This yields $\tau = 83\pm14$~minutes. 

\begin{figure}
    \centering
    \includegraphics{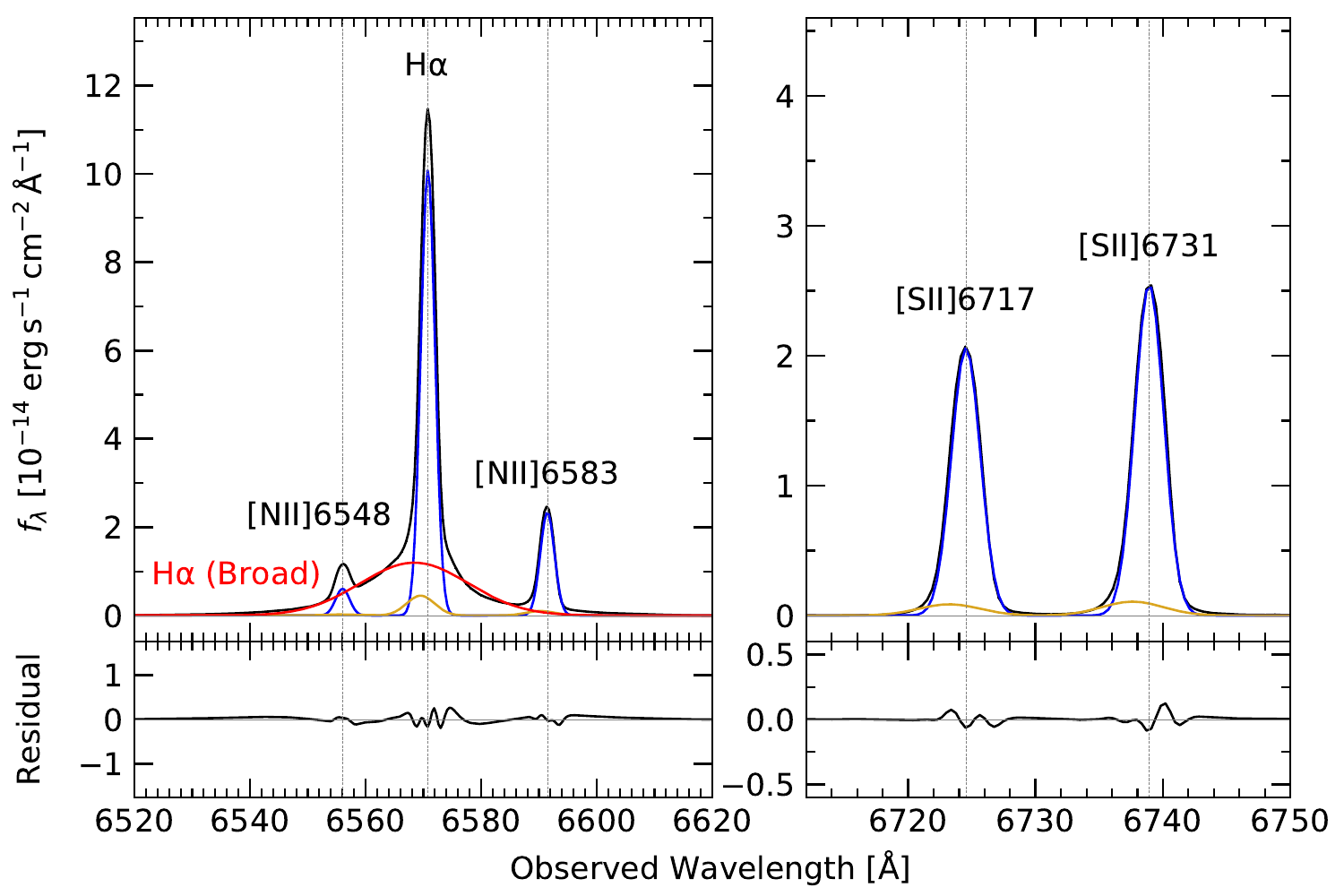}
    \caption{\textbf{Spectral decomposition of the Broad H$\alpha$ emission line region.} The GMOS spectrum was decomposed into the broad H$\alpha$ (red), narrow H$\alpha$, and narrow [NII] components (blue). The width of the broad H$\alpha$ is measured as $\sigma_{\rm line} = 426 \pm 1 \kms$. The narrow lines, which reflect the gravitational potential of the host galaxy 
    and AGN outflows, are fit with a strong core component and a weak broad wing component seen in the isolated [SII] doublet.}
    \label{fig:decomposition}
\end{figure}

The mass of the black hole is given by the virial theorem and can be written $M_{\rm BH} = f (c\tau) \sigma_{\rm line}^2/G$, where $f$ is a scale factor that accounts for the orientation of the BLR to the line-of-sight and anisotropy in the line profile, $\sigma_{\rm line}$ is the line-of-sight velocity dispersion of gas in the BLR, $G$ is the gravitational constant, and $c$ is the speed of light. $\sigma_{\rm line}$ is measured from the width of the broad H$\alpha$ emission line, and we measure $\sigma_{\rm line}= 426\pm1$~km~s$^{-1}$ (see Methods and Figure~\ref{fig:decomposition}). The scale factor is not known, but the typical scale factor can be determined\cite{Onken+04} by matching reverberation masses to their expected values from the empirical correlation between $M_{\rm BH}$ and $\sigma_{\star}$. Using the most updated value\cite{Woo+15}, $\langle f \rangle = 4.47$, we determine $M_{\rm BH} = 9100^{+1500}_{-1600}$ \msun\ in NGC~4395, where the quoted error is calculated based on the measurement uncertainty.

The scale factor includes the projection from the line-of-sight to true velocity dispersion and could be much higher if the broad-line region is viewed almost face-on. There is no direct measurement of the inclination angle, but circumstantial evidence indicates that the mean calibrated value is appropriate to NGC~4395. The mean scale factor and inclination angle among five Seyfert~1 galaxies, which are estimated based on the dynamical modeling with velocity-resolved reverberation measurements, are $f=4.8\pm2.5$ and 20-30~degrees, respectively\cite{Pancoast+15}. 
Meanwhile, the \textit{HST} [OIII] $\lambda$5007\AA\ image shows a biconical outflow in the E-W direction (Methods). Assuming that the outflow is intrinsically axisymmetric, the observed morphology indicates an inclination angle of at least 20~degrees.
The molecular gas disk within about 10~pc of the nucleus is also inclined near 37~degrees\cite{denBrok+15}. We conclude that the scale factor, and thus the mass, is accurate to within a factor of a few. 
The scale factor could be better constrained through velocity-resolved reverberation mapping and dynamical modeling of the broad-line gas. 

Prior Balmer-series measurements\cite{Desroches+06,Edri+12} resulted in less reliable lags, but the velocity measurements (Methods) are roughly consistent with our measurement. When using the same scale factor, we would therefore arrive at a similar $M_{\rm BH}$. Our mass is lower than that measured from a \textit{Hubble Space Telescope} reverberation campaign with the CIV emission line in UV ($M_{\rm BH} = 3.6\pm1.1\times 10^5 M_{\odot}$)\citep{Peterson+05}, but most of the discrepancy results from the choices for line width ($\sigma_{\rm line} \approx 2900$~km~s$^{-1}$, see Methods) and scale factor ($f=5.5$), as their CIV lag measurements in the two \textit{HST} visits ($\tau_1 = 48^{+24}_{-19}$~min and $\tau_2 = 66^{+24}_{-19}$~min) are consistent with ours. 

In contrast, the mass obtained from modeling the nuclear molecular disk\cite{denBrok+15} is $M_{\rm BH}=4^{+8}_{-3}\times 10^5 M_{\odot}$ (3$\sigma$ uncertainties). This measurement relies on separating the central black hole from the nuclear star cluster\cite{Filippenko&Ho03}, which dominates the mass in the region, and is thus not a primary mass measurement. As the authors point out, separating the two gravitational components is complicated by the black hole activity because the accretion disk is a point source with a similar luminosity in the F814W filter ($L_{\rm disk} \approx 1.3\times 10^6 L_{\odot}$) as the nuclear star cluster ($L_{\rm nsc} \approx 1.6\times 10^6 L_{\odot}$)\cite{denBrok+15}. At the \textit{HST}/WFC3 resolution of $\theta \approx 0.1$\arcsec, the angular size of the point source is comparable to the nuclear star cluster ($r_e<0.2$\arcsec). The bright narrow-line region extends to about 0.5\arcsec, and also contributes to the broad-band filters used to measure the mass of the nuclear star cluster. 

\begin{figure}
    \centering
    \includegraphics{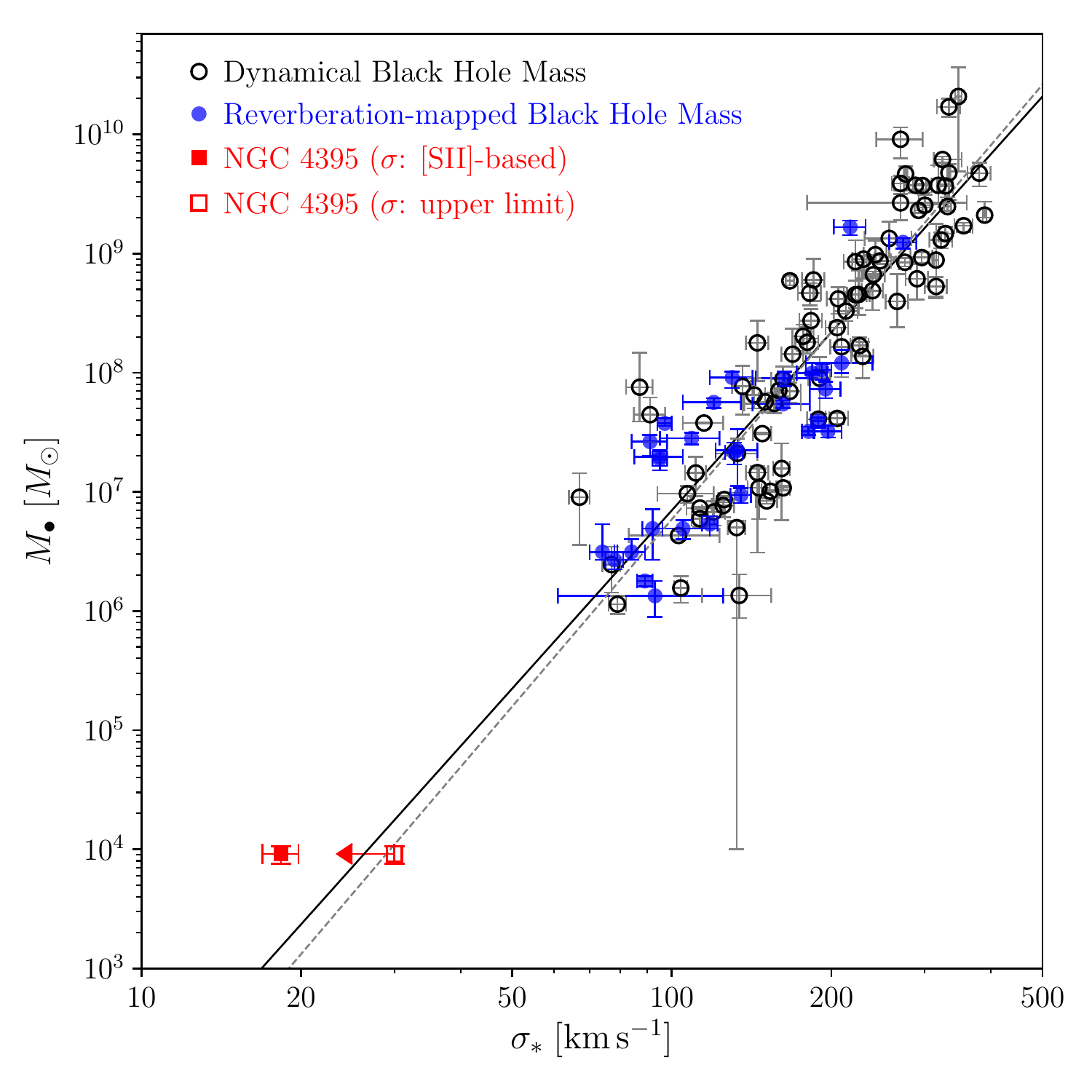}
    \caption{\textbf{$M_{\rm BH} - \sigma_{\star}$ relation.} The $M_{\rm BH} = 9100^{+1500}_{-1600} M_{\odot}$ measurement is plotted in comparison to more massive black holes\cite{Woo+15} by adopting either the 30~km~s$^{-1}$ upper limit to the stellar velocity dispersion\cite{Filippenko&Ho03} [open red square] or the dispersion of the [SII] emission line that we use as a proxy for $\sigma_{\star} = 18\pm1$~km~s$^{-1}$ [filled red square]. 
    Note that the quoted errors are based on the measurement uncertainty.
    Prior measurements are dynamical measurements in inactive galaxies [open black circles] or reverberation mapping in active galaxies [filled blue circles] collected from the previous studies\cite{Woo+15}. The black solid line is the best fit to the combined sample, whereas the black dashed line is the best fit to the galaxies with dynamical mass measurements.
    }
    \label{fig:Msigma}
\end{figure}

Black-hole mass scales with several galaxy properties, most notably the stellar velocity dispersion, $\sigma_{\star}$, of the stellar bulge or pseudo-bulge\cite{Kormendy&Ho13}. The origin of this relation remains controversial, but hierarchical assembly\cite{Jahnke&Maccio11} or AGN feedback\cite{Granato+04,Croton+06} are leading candidates. NGC~4395 has no bulge, but we test whether the $M_{\rm BH}-\sigma_{\star}$ relation extends to the stellar velocity dispersion of the stars around the AGN. We did not clearly detect stellar absorption lines in the GMOS spectrum, so we instead use the [SII] line as a proxy (Methods) and find $\sigma_{\star} = 18\pm1$~km~s$^{-1}$ within 3\arcsec\ of the nucleus, which is consistent with the previous upper limit\cite{Filippenko&Ho03} $\sigma_{\star} < 30$~km~s$^{-1}$. Figure~\ref{fig:Msigma} shows that NGC~4395 is broadly consistent with the extrapolation of the best-fitting relation\cite{Woo+15} to low $\sigma_{\star}$, considering the intrinsic scatter. This indicates that the relationship between $M_{\rm BH}$ and $\sigma_{\star}$ need not depend on hierarchical growth through galaxy mergers that would produce a bulge, and that low-mass galaxies (where feedback from supernovae is more important than that from black holes) need not have under-massive black holes\cite{Dubois+15,Angles-Alcazar+17}.

The incidence of intermediate-mass black holes in low-mass galaxies can determine whether central black holes originate primarily from $100-1000 M_{\odot}$ ``light'' seeds (the remnants of Population~III stars) or $10^4-10^6 M_{\odot}$ ``heavy'' seeds (from the direct collapse of massive gas clouds)\cite{Greene12}. In the latter case, the black hole in NGC~4395 must not have grown appreciably since its formation, and its growth therefore preceded the formation of the nuclear star cluster and small bar. 

Until the advent of 30-m class optical telescopes, reverberation masses will remain the best primary mass measurements for low-mass galaxies beyond the Local Volume. These measurements will establish whether dwarf galaxies generally host intermediate-mass black holes, and how they shape (or are shaped by) their host galaxies. In turn, the fraction of dwarf galaxies hosting black holes may solve the long-standing mystery of how central black holes originally formed. 

\clearpage

\begin{methods}

\section{Observations and data reduction}

\subsection{Spectroscopy.}
On 2017 April 29, we observed the nucleus of NGC~4395 with the GMOS instrument at the 8.1-m Gemini-North telescope. We used the long-slit GMOS mode with the R831 grating (4600-6950\AA) and a 0.75\arcsec\ slit. The spectra were accumulated as a series of 300~s exposures over 3 hours, yielding 36 epochs with a 6~minute cadence. During this time, the seeing degraded from 1\arcsec\ to 2\arcsec. We also obtained a 120~s spectrum of G191-B2B, which is a spectro-photometric standard, using the same instrumental setup. 

We reduced the data with the Gemini pyRAF software package and extracted spectra from the nucleus with an aperture size corresponding to three times the seeing. The instrumental line resolution was measured as $\sigma_{\rm line} = 47$~km~s$^{-1}$ from sky lines. The flux of each spectrum was calibrated using the GMOS spectra from spectrophotometric stars. After doing this, the [SII] line flux still varied between epochs, whereas we expect no variation during the monitoring campaign because it comes from the much larger scale narrow-line region. To create a mean spectrum, we rescaled each spectrum so that the [SII] fluxes matched. 

\subsection{Photometry.}
On 2017 May 2 and 2018 April 8, we observed NGC~4395 in the $V$ and narrow H$\alpha$ bands using several telescopes. Bad weather at Gemini-North precluded simultaneous spectroscopic monitoring for both dates. The narrow H$\alpha$ data were obtained at the MDM 2.4-m telescope in Arizona using the MDM4K camera with 2$\times$2 pixel on-chip binning, resulting in an 11.5~arcmin field of view (FOV) and a pixel scale of 0.34\arcsec/pixel. The H$\alpha$ filter, KP 1468, has a central wavelength of 6567.35\AA, a FWHM of 84.5\AA, and a peak transmission of 71.83\%. We observed in a sequence of several consecutive 300~s H$\alpha$ exposures followed by a single 120~s $V$ band exposure, and repeated this sequence as long as the target was visible. In 2017, we obtained 57 epochs for H\al\ and 19 for $V$, with 1.2\arcsec\ seeing. In 2018, we obtained 80 epochs for H\al\ and 16 for $V$, with 0.9\arcsec\ seeing. We also created flat-field images by observing the twilight sky. 

We simultaneously obtained a continuous series of 300~s $V$ band exposures from the MDM 1.3-m telescope with the Templeton camera (8.5~arcmin FOV and 0.5\arcsec/pixel) and the LOAO 1-m telescope in Arizona with the 4k e2v camera (28.1~arcmin FOV and 0.8\arcsec/pixel). We also observed the source in $V$ at the Mt. Laguna Observatory 1-m telescope in California with the e2V 42-40 2k camera (12.3~arcmin FOV and 0.72\arcsec/pixel after 2$\times$2 binning). The seeing at these sites varied between 0.9-2.5\arcsec. 

The data were reduced following standard procedures for each instrument, and the instrumental magnitudes of the AGN and comparison stars in the field of view were measured using standard aperture photometry. The variability in the AGN was measured from the relative change of the AGN magnitude with respect to those of non-varying comparison stars in the field of view. The uncertainty of the AGN magnitude is derived by combining the measurement uncertainty ($\sim$1\%) and the systematic uncertainty calculated from the standard deviation of the small relative change of comparison stars with respect to the expected magnitudes at each night.

\subsection{Hubble Space Telescope [OIII].}
We used archival images (Proposal ID 12212) from the Wide Field Camera 3 (WFC3) in the F547M and F502N ([OIII] $\lambda$5007\AA) filters, which had exposures of 746~s and 1521~s, respectively. The goal of the HST image analysis was to search for ionized gas outflows in the narrow-line region and constrain the inclination angle of the axis of the AGN outflows to the line-of-sight. The flat-fielded images were drizzled with the AstroDrizzle software and aligned to better than 0.1~pixel. 

\section{Data analysis}

\subsection{Time delay measurement.}
The lag between the $V$ continuum and the H\al\ line emission was measured by cross-correlating the light curves in each band. The cross-correlation coefficient $r$ was calculated for a given time lag $\tau$, ranging from -4 to 4 hours, using the interpolated cross-correlation function (ICCF)\cite{White&Peterson94}. The peak $r$ in this range corresponds to the most likely lag, and can either be defined as the maximum value or as the centroid of the $r$ distribution, corresponding to $\tau_{\rm peak}$ and $\tau_{\rm cent}$, respectively (Extended Data Figure~\ref{fig:lagdist}). To estimate the true values and their corresponding uncertainties, we measured the centroid and peak $r$ in each of 1,000 realizations of the light curve based on randomly selected subsets of the data, of which fluxes were randomized according to their measurement uncertainties \cite{Peterson+98}. The median value and central 68\% interval of the distribution are adopted as the time lag and and its uncertainty, respectively. This procedure was applied to the measured light curves and yields centrally peaked, approximately symmetric $r$ distributions. We thus adopt $\tau_{\rm cent}= 55^{+28}_{-32}$~min and $\tau_{\rm cent} = 49^{+14}_{-15}$~min for 2017 May 2 and 2018 April 8, respectively. 

These values are lower limits to the true time delay because the H$\alpha$ filter contains both the variable H$\alpha$ line as well as continuum from the accretion disk. The mean continuum contribution is estimated at 18.3\% by convolving the transmission curve of the MDM H$\alpha$ filter with the mean GMOS spectrum from 2017 April 29, where the contribution of each component is resolved. When using single-epoch GMOS spectra, the contribution varies by 0.3\% (1$\sigma$). The continuum fraction in each H$\alpha$ epoch is then estimated by assuming that the continuum flux variability in the $H\alpha$ band (6450--6650\AA) is the same as in $V$ (4800--6500\AA) and scaling the fractional $V$-band variability by multiplying a factor of 0.183.
The lag measured after subtracting this flux in each H$\alpha$ epoch is $\tau_{\rm cent} = 83\pm14$~min for the higher quality 2018 April 8 data set, which we adopt as the true lag. Note that the variability of the continuum in the H$\alpha$ filter is expected to be no greater than that of the continuum in the V filter. Thus, this correction is a maximum correction if the variability amplitude anti-correlates with the observed wavelength of AGN continuum. Considering the close wavelength ranges of the V and H$\alpha$ bands, we expect that the continuum variability in the two filters is similar. The H$\alpha$ lag measurement can be improved by simultaneous spectroscopic and photometric monitoring, since the decomposition of the broad H$\alpha$ line from AGN continuum at each epoch will provide a more reliable measurement of the lag. 

\subsection{Photo-ionized gas velocity.}
We measure the gas velocity by fitting the line profile in the mean GMOS spectrum to obtain the velocity dispersion (Figure~3). We subtract the featureless continuum and simultaneously fit the narrow H$\alpha$ and [NII] doublet ($\lambda\lambda$6548,6583\AA) components with the broad H$\alpha$ line. The nearby, isolated [SII] line doublet ($\lambda\lambda$6717,6731\AA) has both a strong core and a small blueshifted wing, so we fit the narrow [NII] and H$\alpha$ lines with two Gaussian models in which the flux ratio between the core and wings is fixed at the value measured for [SII]. The dispersion of the core and wing components for [SII] are $\sigma_{\rm core} = 18\pm1$~km~s$^{-1}$ and $\sigma_{\rm wing} = 100\pm5$~km~s$^{-1}$, indicating that the blue wing is an outflow of ionized gas. The dispersion of the core reflects the gravitational potential in the central 100~pc. The broad line is fit with a single Gaussian model with $\sigma = 426\pm1$~km~s$^{-1}$ (full-width at half-maximum FWHM$=1003\pm3$~km~s$^{-1}$). This is roughly consistent with prior measurements in several lines, including FWHM$=1500\pm500$~km~s$^{-1}$ in H$\alpha$\cite{Kraemer+99}, FWHM$=1175\pm325$~km~s$^{-1}$ in H$\beta$\cite{Edri+12}, FWHM$=755\pm7$ in Pa$\beta$\cite{LaFranca+15}, and FWHM$=1200-1500$~km~s$^{-1}$ in CIV\cite{Peterson+05}.

Prior reverberation mapping studies have used different quantities for the velocity, which accounts for much of the range of masses in the literature. These include the line dispersion (i.e., the second moment of the line profile) or FWHM measured from the variable part of the line, as estimated from the root-mean square spectrum instead of the mean\cite{Peterson+05}, and the line dispersion directly meaasured from the observed line profile instead of using a fitted line profile (e.g., a Gaussian or Lorentzian model). The high quality GMOS data demonstrate that the broad H$\alpha$ line is approximately symmetric and clearly separable from the narrow lines, and we find that these approaches give consistent answers. For example, the rms spectrum constructed from 33 GMOS exposures (three were not used because of cosmic rays coincident with the H$\alpha$ line) has a dispersion of $\approx$400~km~s$^{-1}$, which is similar to the 426~km~s$^{-1}$ measured from the mean spectrum. As the rms spectrum represents the variable part of the emission line, the line width can be different between the rms and mean spectra, if a part of the line does not vary. In the case of CIV, for example, both the line dispersion and FWHM are reported as $\sim$3000~km~s$^{-1}$ in the previous \textit{HST} reverberation campaign\cite{Peterson+05}, which is a factor of $\sim$2 larger than the line dispersion and FWHM measured from the mean spectrum. CIV and H$\alpha$ do not have the same line profile and the characteristics of the varying component in CIV seems different from that of H$\alpha$. In addition, the rms spectrum is more susceptible to bias towards higher line dispersion than the mean spectrum since the relatively low variability makes much weaker signals in the rms spectrum, particularly at the wing. 

If we measure the CIV line width from the mean spectra in the same way as for H$\alpha$, we obtain $\sigma_{\rm CIV} \approx 1500$~km~s$^{-1}$. For the same scale factor $f=4.47$, the discrepancy between the Balmer-series and CIV-based masses is a factor of $\sim 10$. Note that the line profile of CIV is far from a Gaussian profile as the line dispersion is larger than FWHM (see Table 4 in the previous study\cite{Peterson+05}). As the FWHM of CIV measured from the mean spectra ($\sim$1200-1500~km~s$^{-1}$) is comparable to that of H$\alpha$, we obtain CIV reverberation mass as $\sim$2 $\times$ 10$^4$ \msun, if we use the CIV FWHM and the corresponding scale factor f=1.12 from the most recent calibration\cite{Woo+15}. This mass is only a factor of 2 larger than H$\alpha$ reverberation mass. Therefore, the choice of the velocity measure yields a large range of the CIV reverberation mass, while we obtain a consistent reverberation mass from of H$\alpha$, using either line dispersion or FWHM measured from mean or rms spectrum. 

\subsection{Inclination angle of [OIII] outflows.} If we assume that ionized gas outflows are parallel to the axis of the accretion disk, the morphology of the [OIII] emission contains information about the line-of-sight inclination angle of the accretion disk. The \textit{HST} image obtained with the narrow [OIII] filter F502N shows an asymmetric distribution around the central point source (Extended Data Figure~\ref{fig:OIII}). A comparison of the F502N image to the nearby continuum image obtained with the medium filter F547M indicates that this structure is not a part of the point-spread function, and is most likely biconical ionized gas outflows. In order to remove the continuum in the F502N filter, we subtract the scaled F547M filter image from the F502N filter image, by matching the flux of the central point source (see right panel in Extended Data Figure~\ref{fig:OIII}). In contrast to the expected symmetric [OIII] flux distribution for the case of a face-on disk, the asymmetric [OIII] morphology indicates that the line-of-sight inclination angle is much greater than zero. Although the inclination angle is not directly measurable from the [OIII] image, the biconical morphology suggests that it should be close to the average inclination of 20 degrees for type 1 AGNs, which was measured based on dynamical modeling and velocity-resolved lag measurements\cite{Pancoast+15}. 

\subsection{Stellar velocity dispersion proxy.}
The narrow emission lines come from gas that is ionized primarily by the AGN but whose velocity is determined by the gravitational potential within the central $\sim$100~pc. Since the gas mass is much smaller than that of the stars, the velocity dispersion of this gas should reflect that in the stars. We measured the velocity dispersion of the narrow H$\alpha$ line, but as it is blended with the broad component, the isolated [SII] doublet provides a better measurement.  We fit the lines with two components as described above, and $\sigma_{\rm line} = 18\pm1$~km~s$^{-1}$ for the dominant core. This is consistent with the value measured for H$\alpha$. Note that the kinematics of low-ionization line gas typically follows that of stars, showing a good correlation between the velocity dispersion of gas and stars\cite{Komossa+08}. Thus, we use the velocity dispersion of gas as a proxy for the stellar velocity dispersion.

\subsection{Data availability.} The data on which this study was based are available from the corresponding author upon reasonable request. 

\newpage
\setcounter{figure}{0} 

\begin{figure}
    \centering
    \includegraphics{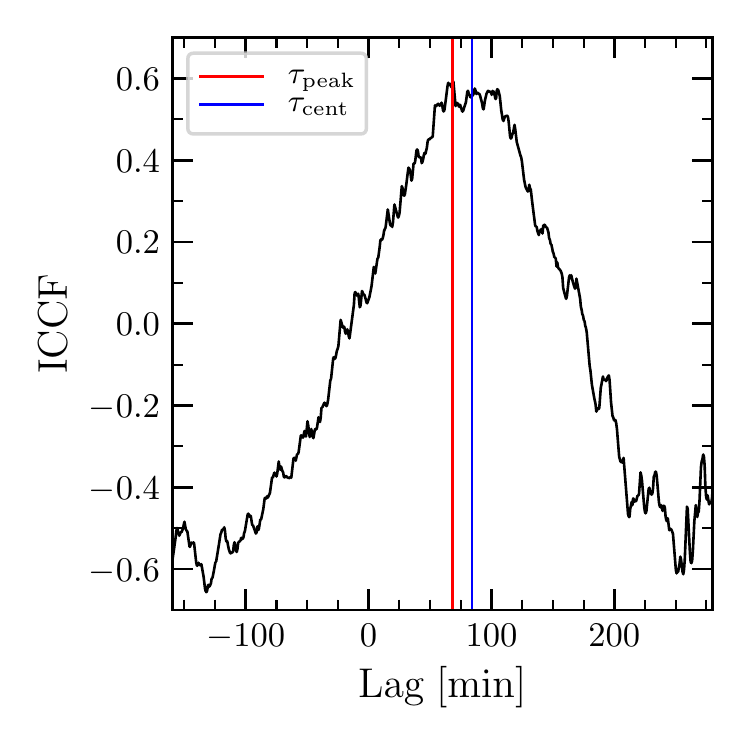}
    \includegraphics{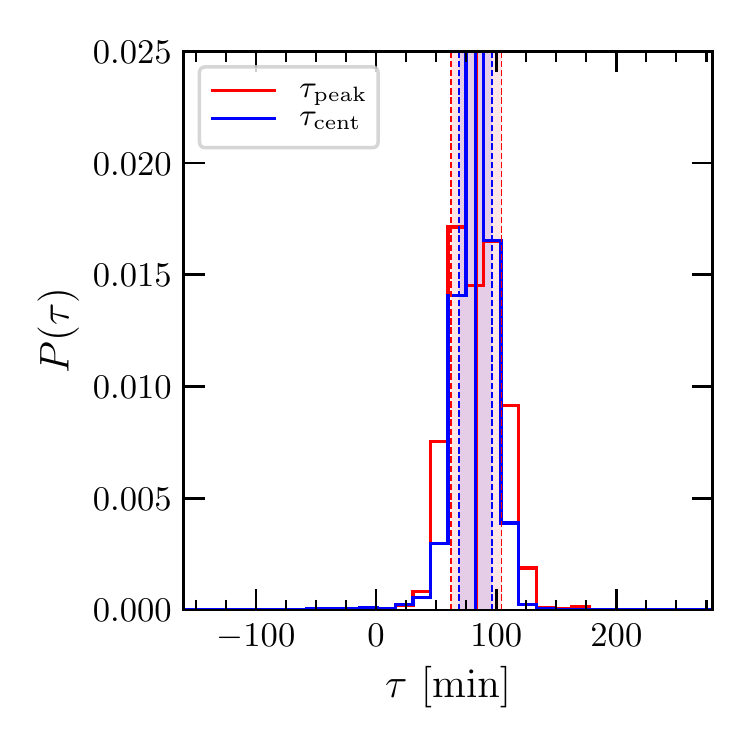}
    \caption{\textbf{Cross-correlation $V - $H$\alpha$ lag measurements.} The peak and centroid lags for the cross-correlation coefficient are denoted with red and blue lines, respectively (left panel). We define the centroid time lag ($\tau_{\rm cent}$) as the ICCF-weighted average lag for values larger than 80\% of the maximum ICCF.
    We adopt the 
    median and central 68\% interval
    width $\tau_{\rm cent} = 83\pm14$~min from the distribution of $\tau_{\rm cent}$ as measured from 1000 realizations of the $V$ and H$\alpha$ light curves. The distributions of $\tau_{\rm peak}$ and $\tau_{\rm cent}$ are denoted with red and blue lines, respectively, in the right panel. }
    
    \label{fig:lagdist}
\end{figure}

\begin{figure}
    \centering
    \includegraphics[width=\textwidth]{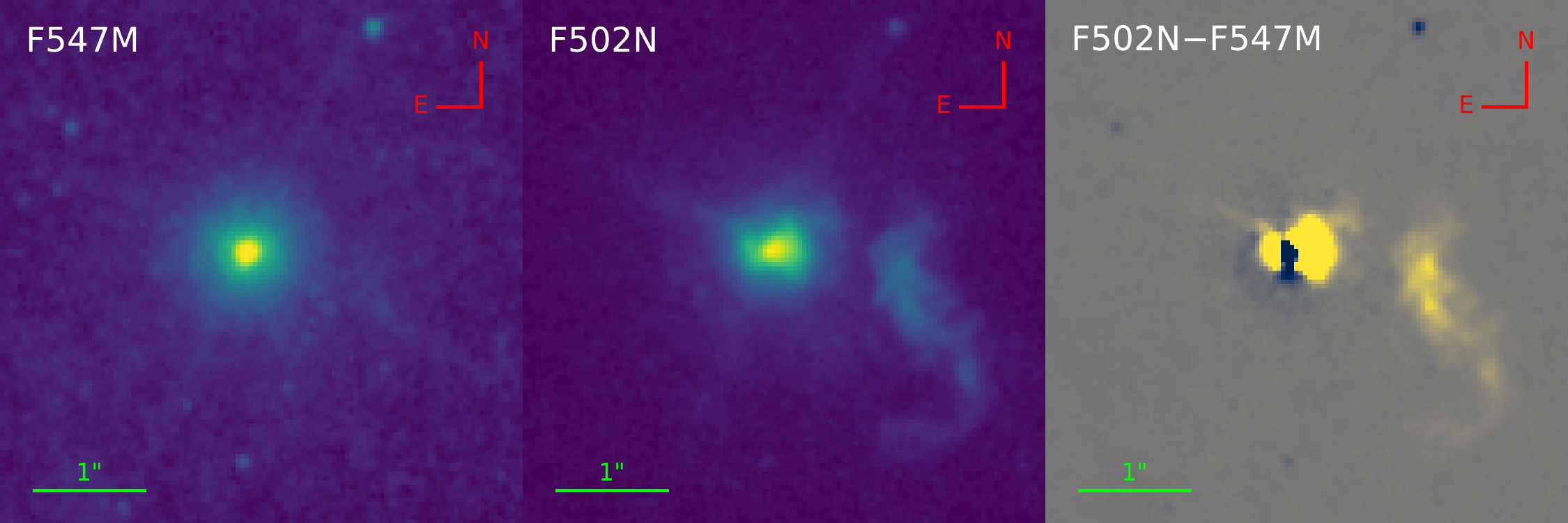}
    \caption{\textbf{Biconical [OIII] outflow.} The \textit{HST} F547M continuum filter (left) shows the stellar continuum from the nuclear star cluster and the continuum emission from the AGN. The F502N filter image (center) shows the [OIII] emission from ionized gas, but includes some continuum emission. After subtracting the scaled continuum from the F502N image, only the [OIII] remains (right). The continuum-subtracted image reveals a biconical, but asymmetric, outflow, which indicates that the axis of the [OIII] outflow is inclined relative to our line of sight.
    }
    \label{fig:OIII}
\end{figure}

\clearpage

\end{methods}



\bibliographystyle{naturemag}


\begin{addendum}
 \item This work has been supported by the Basic Science Research Program through 
the National Research Foundation of Korea government (2016R1A2B3011457). 
We thank the various contribution from the NGC 4395 Collaboration.
 \item[Author contributions] J.W. wrote the manuscript with comments from all authors and performed much of the analysis. J.W. also carried out the Gemini observations and coordinated all observations. H.C. analyzed the photometric light curves and Gemini spectra to measure the lag and velocity dispersion. E.G. substantially revised the manuscript and contributed to the analysis. E.H-K. carried out MDM observations and revised the manuscript. H.A.L., J.S., and D.S. carried out MDM observations. J.H. carried out MLO observations.
 \item[Competing Interests] The authors declare that they have no
competing financial interests.
 \item[Correspondence] Correspondence and requests for materials
should be addressed to J.W.~(email: woo@astr.snu.ac.kr).
\end{addendum}



\end{document}